**Title**

# Robust frequency-dependent diffusion kurtosis computation using an efficient direction scheme, axisymmetric modelling, and spatial regularization

**Authors & Affiliations**

Jake Hamilton[1,2], Kathy Xu[3,4], Arthur Brown[3,4], Corey A. Baron[1,2]

1. Centre for Functional and Metabolic Mapping (CFMM), Robarts Research Institute, University of Western Ontario, London, Ontario, Canada
2. Department of Medical Biophysics, Schulich School of Medicine and Dentistry, University of Western Ontario, London, Ontario, Canada
3. Translational Neuroscience Group, Robarts Research Institute, Schulich School of Medicine and Dentistry, University of Western Ontario, London, Ontario, Canada
4. Department of Anatomy and Cell Biology, University of Western Ontario, London, Ontario, Canada

**Corresponding author:** Corey A. Baron (corey.baron@uwo.ca)

**Abstract**

Frequency-dependent diffusion MRI (dMRI) using oscillating gradient encoding and diffusion kurtosis imaging (DKI) techniques have been shown to provide additional insight into tissue microstructure compared to conventional dMRI. However, a technical challenge when combining these techniques is that the generation of the large b-values required for DKI is difficult when using oscillating gradient diffusion encoding. While efficient encoding schemes can enable larger b-values by maximizing multiple gradient channels simultaneously, they do not have sufficient directions to enable fitting of the full kurtosis tensor. Accordingly, we investigate a DKI fitting algorithm that combines axisymmetric DKI fitting, a prior that enforces the same axis of symmetry for all oscillating gradient frequencies, and spatial regularization, which together enable robust DKI fitting for a 10-direction scheme that offers double the b-value compared to traditional direction schemes. Using data from mice (oscillating frequencies of 0, 60, and 120 Hz) and humans (0 Hz only), we first show that axisymmetric modelling is advantageous over full kurtosis tensor fitting in terms of preserving contrast and reducing noise in DKI maps, and improved DKI map quality when using an efficient encoding scheme with averaging as compared to a traditional scheme with more encoding directions. We also demonstrate how spatial regularization during fitting preserves spatial features better than using Gaussian filtering prior to fitting, which is an oft-reported preprocessing step for DKI, and that enforcing consistent axes of symmetries across frequencies improves fitting quality. Thus, the use of an efficient 10-direction scheme combined with the proposed DKI fitting algorithm provides robust maps of frequency-dependent directional kurtosis parameters that can be used to explore novel biomarkers for various pathologies.

## 1. Introduction

Diffusion MRI (dMRI) is a well-known technique that allows probing of tissue microstructure on spatial scales unattainable with conventional MRI techniques. Diffusion kurtosis imaging (DKI) is a form of dMRI that quantifies the non-Gaussian component of diffusion within tissue, providing greater sensitivity to microstructural changes as compared to conventional diffusion tensor imaging (DTI) (Jensen et al., 2005; Jensen & Helpern, 2010). DKI has been used to study various pathologies such as stroke (Grinberg et al., 2012; Lampinen et al., 2021), mild traumatic brain injury (Stenberg et al., 2021; Stokum et al., 2015), neurodegenerative diseases (Chu et al., 2022; Dong et al., 2020; Falangola et al., 2013; Vanhoutte et al., 2013), as well as many others (Steven et al., 2014), and has shown increased sensitivity to changes in tissue microstructure than DTI. An important sequence parameter to consider when interpreting changes in dMRI metrics is the effective diffusion time ($\Delta t_{eff}$) which determines the length scale within the tissue being probed. Commonly used pulsed gradient spin echo (PGSE) sequences to encode diffusion are limited to relatively long $\Delta t_{eff}$ (> 10 ms) for typical pre-clinical gradient hardware (Jelescu et al., 2022; Portnoy et al., 2013), meaning the acquired signal reflects diffusion restriction at multiple spatial scales (i.e., both cellular and sub-cellular). Conversely, encoding diffusion using an oscillating gradient spin echo (OGSE) sequence allows much shorter $\Delta t_{eff}$ (corresponding to higher OGSE frequencies), increasing sensitivity to smaller spatial scales (Schachter et al., 2000). By collecting data at multiple gradient oscillation frequencies (i.e., frequency-dependent dMRI), the spatial scale sensitivity can be varied, enabling investigation of tissue microstructure at both sub-cellular and cellular levels, increasing sensitivity and specificity for studying various pathological conditions. Exploiting the frequency-dependence of DTI metrics has been well-characterized in the healthy brain (Baron & Beaulieu, 2014), as well as in the study of stroke (Baron et al., 2015), cancer (Iima et al., 2019; Maekawa et al., 2020), and in neurodegenerative disease (Aggarwal et al., 2014). Conversely, the dependence of DKI metrics on OGSE frequency is less understood but has been characterized in healthy humans (Dai et al., 2023) and rodents (Aggarwal et al., 2020; Pyatigorskaya et al., 2014), as well as in rodent studies of hypoxic-ischemic injury (Wu et al., 2018) and a demyelination disease model (Aggarwal et al., 2020).

A key technical challenge in using frequency-dependent DKI is that OGSE sequences are much less efficient at generating the large b-values required for DKI compared to PGSE, as b-value is related to the gradient strength (G) and OGSE frequency (f) through $b{\sim}G^2/f^3$ (Xu, 2021). Consequently, tetrahedral (4 directions) encoding schemes (Conturo et al., 1999) have been required in humans (Borsos et al., 2023) and mice (Aggarwal et al., 2020) to generate the desired b-value for diffusion kurtosis fitting. However, this direction scheme does not allow for directional parameters to be calculated and likely introduces rotational variance in computed metrics (Jones, 2004; Nilsson et al., 2020). Directional (axial and radial) kurtosis parameters have been shown to provide additional specificity to changes in mean kurtosis (Conklin et al., 2016; Goryawala et al., 2018), but estimation of the full kurtosis tensor requires at least two non-zero b-value shells with 15 independent diffusion directions (Lu et al., 2006). A recently developed fitting method, axisymmetric DKI (Hansen et al., 2016), allows fitting of directional kurtosis parameters with two non-zero b-value shells and 9 independent diffusion directions. This method requires determination of a symmetric axis of diffusion within each voxel to which kurtosis parameters are fit to based on the relation between encoding direction and the symmetric axis. This method has been shown to generate comparable and even improved kurtosis maps to those computed via kurtosis tensor fitting (Hansen et al., 2016; Jespersen, 2018; Oeschger et al., 2023).

A second challenge when using DKI is noise propagation, primarily due to the high diffusion-weighting that is required to compute kurtosis in addition to the large number of parameters to be fit when estimating the kurtosis tensor (i.e., noise amplification) (Glenn et al., 2015; Tabesh et al., 2011). Subsequently, Gaussian smoothing on diffusion-weighted images prior to fitting is widely accepted as a necessary pre-processing step used to reduce noise levels in kurtosis maps (Jensen et al., 2005; Tax et al., 2022). However, because there is averaging of signal from neighboring voxels, this leads to blurring around sharp edges within the image and can cause bias in quantitative analysis (Falconer & Narayana, 1997; Pfefferbaum & Sullivan, 2003; Vos et al., 2011). An alternative approach is to use spatial regularization (Tikhonov et al., 1995) during fitting which balances smoothing of noise with the fitting residual. Regularization has been shown to be advantageous over Gaussian smoothing in functional MRI studies in terms of retaining original image contrast and detection of true areas of activation (Casanova et al., 2009; Liu et al., 2010; Ou & Golland, 2005). It has also been used to control noise amplification in various avenues of dMRI (McGraw et al., 2009; Veraart et al., 2016; Wu & Yan, 2021), however, its use during DKI fitting remains underutilized.

In this study we aim to delineate an acquisition and analysis scheme capable of generating high quality frequency-dependent kurtosis maps. We present a 10-direction scheme that is two times more efficient in generating b-value compared to traditional schemes, mitigating a key technical challenge when combining OGSE and DKI techniques. Further, we propose that using data from all frequencies and b-values to estimate the symmetric axis of diffusion will provide a more robust estimate that will reduce noise in parameter maps. We then introduce a two-step spatial regularization algorithm to reduce noise both during symmetric axis determination and kurtosis fitting. We also provide direct comparison demonstrating the advantage of collapsing data from all frequencies to determine the axis of symmetry and using this regularization algorithm over conventional Gaussian smoothing.

## 2. Methods

### *2.1. Human connectome project (HCP) data*

Pre-processed dMRI data of a healthy female subject from the HCP1200 release of the Human Connectome Project (HCP) (McNab et al., 2013; Sotiropoulos et al., 2013) was used for validation of axisymmetric DKI fitting and spatial regularization. dMRI data was acquired on a 3 Tesla (T) scanner with parameters: 1.25 x 1.25 x 1.25 $mm^3$ voxel size, TE/TR = 89.5/5520 ms. Data consisted of 18 b=0 images and 3 b-value shells of 1000, 2000, and 3000 $s/mm^2$, with 90 directions at each. Data from the highest b-value shell (3000 $s/mm^2$) was excluded as the kurtosis fit is known to be most robust at b-values $\leq$ 2500 s/mm (Jelescu et al., 2022; Kiselev & Il'yasov, 2007). Full details regarding data acquisition can be found in the HCP 1200 subjects reference manual (https://www.humanconnectome.org/study/hcp-young-adult/document/1200-subjects-data-release).

*2.2. Subject info*

Data was collected from 8 transgenic mice (4 males) with humanized *mapt* and *app* genes knock in at 6 months of age as part of another ongoing study, and one female mouse of the same genotype and age to compare our efficient direction scheme with a traditional scheme. All animal procedures were approved by the University of Western Ontario Animal Care Committee and were consistent with guidelines established by the Canadian Council on Animal Care. Before scanning sessions, anesthesia was induced by placing mice in an induction chamber with 4% isoflurane and an oxygen flow rate of 1.5 L/min. Throughout the scanning session, isoflurane was maintained at 1.8% with an oxygen flow rate of 1.5 L/min through a custom-built nose cone.

*2.3. Data acquisition & pre-processing*

In vivo MR scanning sessions were performed on a 9.4 T Bruker small animal scanner equipped with a gradient coil insert of 1 T/m strength (slew rate = 4,000 T/m/s). Both anatomical and diffusion scans were acquired for each subject. Anatomical images were acquired using a T2-weighted TurboRARE sequence with parameters: in-plane resolution 100 x 100 $\mu m^2$, slice thickness 500 µm, TE/TR = 30/5500 ms, 24 averages, scan time of 25 minutes. The dMRI protocol included a PGSE sequence (gradient duration = 9.4 ms and diffusion time = 12.4 ms) and OGSE sequences with frequencies of 60 and 120 Hz (corresponding $\Delta t_{eff}$ of 4.2 and 2.1 ms (Does et al., 2003)). The lowest OGSE frequency (60 Hz) implements the recently introduced frequency tuned bipolar waveforms to reduce the TE of the acquisition (Borsos et al., 2023). For all frequencies, data consisted of 2 b = 0 images and two b-value shells of 1000 and 2500 $s/mm^2$ each with the 10-direction scheme outlined in Table 1. This scheme is an efficient 6-direction scheme (Baron & Beaulieu, 2014) combined with a tetrahedral scheme (Aggarwal et al., 2020; Borsos et al., 2023), which offers a factor of 2 larger b-value than typical schemes due to at least two gradient channels being simultaneously at maximum for each direction. The dMRI protocol was acquired in one integrated scan using single-shot echo planar imaging with 80% of k-space being sampled in the phase encode direction and parameters: in-plane resolution 200 x 200 $\mu m^2$, slice thickness 500 µm, TE/TR = 35.5/15000 ms, 4 averages, total scan time of 66 minutes. To compare our direction scheme with a traditional scheme, two diffusion protocols were acquired during a single scanning session in one mouse. Both protocols had the same scan time of 66 minutes and OGSE frequencies of 0, 60, and 120 Hz. The first protocol was the same as described above: TE/TR = 35.5/15000 ms, 10 directions (Table 1), 4 averages, 4 b=0 images, and the second (traditional) protocol with parameters: TE/TR = 52/15000 ms, 40 directions isotopically distributed using electrostatic repulsion (Jones et al., 1999), 1 average, 4 b=0 images. Note the much longer TE required for the 2nd protocol in order to achieve the same b-value with the less efficient direction scheme. For all acquisitions, complex-valued averages were combined using in-house MATLAB code which included partial Fourier reconstruction, phase alignment, frequency and signal drift correction, and Marchenko-Pastur denoising (Veraart et al., 2016), similar to earlier work (Rahman et al., 2021). Data then underwent Gibbs ringing correction using MRtrix3 (Tournier et al., 2019) followed by EDDY (Andersson & Sotiropoulos, 2016) from FMRIB Software Library (FSL, Oxford, UK) (Smith et al., 2004) to correct for eddy current induced distortions.

| **x** | **y** | **z** |
|---|---|---|
| $0$ | $1$ | $1$ |
| $0$ | $1$ | $-1$ |
| $1$ | $0$ | $1$ |
| $1$ | $0$ | $-1$ |
| $1$ | $1$ | $0$ |
| $1$ | $-1$ | $0$ |
| $\sqrt{2/3}$ | $\sqrt{2/3}$ | $\sqrt{2/3}$ |
| $\sqrt{2/3}$ | $\sqrt{2/3}$ | $-\sqrt{2/3}$ |
| $\sqrt{2/3}$ | $-\sqrt{2/3}$ | $\sqrt{2/3}$ |
| $-\sqrt{2/3}$ | $\sqrt{2/3}$ | $\sqrt{2/3}$ |

***Table 1.*** *The 10-direction scheme.*

*2.4. Data fitting*

*2.4.1. Axisymmetric DKI*

The representation of the diffusion-weighted signal in DKI is given by (Jensen et al., 2005):

$$\log\left(\frac{S_{b,\hat{n}}}{S_0}\right) = -bD_{\hat{n}} + \frac{b^2}{6}\bar{D}^2 W_{\hat{n}} \tag{1}$$

where $S_{b,\hat{n}}$ is the diffusion-weighted signal at a particular diffusion weighting, b, and encoding direction $\hat{n}$, $S_0$ is the signal intensity at b=0, D is the diffusion tensor, and W is the kurtosis tensor. As outlined by Hansen et al. (2016), in systems assumed to have axial symmetry (i.e., in the diffusion tensor, $\lambda_2 = \lambda_3$), kurtosis is characterized by three independent parameters: mean kurtosis ($\overline{W}$), radial kurtosis ($K_\perp$), and axial kurtosis ($K_{||}$). In such a system, the elements of W and D measured along any diffusion direction are given by:

$$W(\theta) = \frac{1}{16}\left(\cos(4\theta)\left(10W_\perp + 5W_\parallel - 15\overline{W}\right) + 8\cos(2\theta)\left(W_\parallel - W_\perp\right) - 2W_\perp + 3W_\parallel + 15\overline{W}\right) \tag{2}$$

and

$$D(\theta) = D_\perp + \cos^2(\theta)\left(D_\parallel - D_\perp\right) \tag{3}$$

where $\theta$ is the polar angle between the axis of symmetry and the diffusion encoding direction, $D_\perp$ is the radial diffusivity, and $D_\parallel$ is the axial diffusivity. Subsequently, mean diffusivity ($\overline{D}$), fractional anisotropy (FA), $K_\parallel$ and $K_\perp$ can be calculated as:

$$\overline{D} = \frac{2D_\perp}{3} + \frac{D_\parallel}{3} \quad FA = \sqrt{\frac{3}{2}\frac{(D_\parallel - \overline{D})^2 + 2(D_\perp - \overline{D})^2}{D_\parallel^2 + 2D_\perp^2}} \quad K_\parallel = \frac{W_\parallel^2 \cdot \overline{D}^2}{D_\parallel^2} \quad K_\perp = \frac{W_\perp^2 \cdot \overline{D}^2}{D_\perp^2} \tag{4,5,6,7}$$

### *2.4.2. Spatial regularization*

For maps with spatial regularization, a two-step algorithm was used to: (1) provide a robust estimate of the symmetric axis in each voxel, and (2) reduce noise amplification in parameter maps during the fitting process. Regularization was implemented using the conjugate gradient method (Ginsburg, 1963), solved with ordinary least squares optimization. Ordinary least squares was chosen for all optimizations as this method is known to have reduced bias relative to weighted least squares (Morez et al., 2023).

*Step One:* In the first step, fitting of the diffusion tensor was regularized for the purpose of determining the axis of symmetry within each voxel, using isotropic total variation (Rudin et al., 1992), where $\gamma_{DT}$ controls the strength of regularization:

$$argmin\|A_{DTI}x_{DT} - y\|_2^2 + \gamma_{DT}\|T_{DT}x_{DT}\|_2^2 \quad (8)$$

The data consistency term is based on the diffusion tensor $x_{DT}$, the encoding matrix for the diffusion tensor representation $A_{DTI}$, and the log-transformed signal data y. The n'th row of $A_{DTI}$ is given by: $A_{DTI,n} = (1, -b_{xx,n}, -b_{yy,n}, -b_{zz,n}, -2b_{xy,n}, -2b_{xz,n}, -2b_{yz,n})$ where n counts the diffusion directions acquired, and for the m'th voxel, $x_{DT,m} = (\log(S_{0,m}), D_{xx,m}, D_{yy,m}, D_{zz,m}, D_{xy,m}, D_{xz,m}, D_{yz,m})$, where $D_{ij}$ are components of the symmetric diffusion tensor (i.e., $x_{DT}$ is a vector of length 7N, where N is the total number of voxels). For each voxel position m, the operator $T_{DT,m}$ performs a numerical derivative, $d_i$, along each of the three spatial dimensions, for each diffusion tensor component:

$$\mathrm{T}_{\mathrm{DT,m}}x_{DT,m} = \left(w_{xx}d_xD_{xx,m}, w_{xx}d_yD_{xx,m}, w_{xx}d_zD_{xx,m}, w_{yy}d_xD_{yy,m}, \dots, w_{yz}d_zD_{yz,m}\right) \quad (9)$$

where $w_{ij}$ is a constant that scales results. For all cases here, $w_{xx} = w_{yy} = w_{zz} = 1$ and $w_{xy} = w_{xz} = w_{yz} = 2$ to account for diffusion tensor cross terms typically having smaller magnitudes than the diagonal of the tensor. The total size of $\mathrm{T_{DT}}x_{DT}$ is 18N, which corresponds to three derivatives for each of the six unique elements of the diffusion tensor.

*Step Two:* In the second step, we regularize the axisymmetric DKI fitting for the purpose of controlling noise amplification in parameter maps using:

$$argmin\|A_{DKI}x_{DK} - y\|_2^2 + \gamma_{DK}\|T_{DK}x_{DK}\|_2^2 \tag{10}$$

where $x_{DK}$ are the diffusion kurtosis parameters, $A_{DKI}$ is the encoding matrix for the axisymmetric DKI model, and $\gamma_{DK}$ is the regularization weighting for this step. The n'th row of $A_{DKI}$ is given by (Eq. 2):

$$A_{DKI,n} = \begin{pmatrix} 1, -b_n(1 - cos^2\theta_n), -b_n(cos^2\theta_n), \frac{{b_n}^2}{6}\left(\frac{10}{16}\cos(4\theta_n) - \frac{8}{16}\cos(2\theta_n) - \frac{2}{16}\right), \\ \frac{{b_n}^2}{6}\left(\frac{5}{16}\cos(4\theta_n) + \frac{8}{16}\cos(2\theta_n) + \frac{3}{16}\right), \frac{{b_n}^2}{6}\left(\frac{-15}{16}\cos(4\theta_n) + \frac{15}{16}\right) \end{pmatrix} \tag{11}$$

where $\theta_n$ is determined from the dot product of the diffusion encoding vector with the symmetric axis of diffusion within each voxel (primary eigenvector of the fitted diffusion tensor from step one), and for the m'th voxel, $x_{DK,m} = (\log(S_{0,m}), D_{\perp,m}, D_{\|,m}, \overline{D}^2 W_{\perp,m}, \overline{D}^2 W_{\|,m}, \overline{D}^2 \overline{W}_m)$. Similar to step one, the operator $T_{DK,m}$ performs a numerical derivative in every voxel along each of the three spatial dimensions, for each diffusion parameter in $x_{DK,m}$:

$$\mathrm{T_{DK,m}}x_{DK,m} = \left(d_x D_{\perp,m}, d_y D_{\perp,m}, d_z D_{\perp,m}, d_x D_{\|,m}, \ldots, d_z \overline{D}^2 \overline{W}_m\right) \tag{12}$$

which has a total size of 15N. Units of $\mathrm{ms/\mu m^2}$ are used for b-values in all calculations to ensure that all the entries of $\mathrm{T_{DK,m}}x_{DK,m}$ are of comparable magnitude to each other since typically b ~ 1 $\mathrm{ms/\mu m^2}$.

A 'base regularization' (mouse data: $\gamma_{DT} = 0.5$, $\gamma_{DK} = 0.075$, human data: $\gamma_{DT} = 0.5$, $\gamma_{DK} = 0.2$) was heuristically chosen such that regularization at both steps contributed to the reduction of noise in subsequent parameter maps while avoiding over-regularization. To investigate the effect of varying the net regularization, the base regularization weightings were multiplied by a single factor.

Our implementation of axisymmetric fitting with optional spatial regularization is available in the MatMRI toolbox at https://gitlab.com/cfmm/matlab/matmri (Baron, 2021; Varela-Mattatall et al., 2023).

### *2.5. Data Analysis*

#### *2.5.1. Comparing kurtosis tensor vs. axisymmetric DKI fitting*

The DIPY project (Henriques et al., 2021) was used to compute kurtosis metrics via kurtosis tensor fitting to compare with the axisymmetric fitting method. Ordinary least squares was used for fitting the kurtosis tensor and metrics were computed using the analytical solution as outlined previously (Tabesh et al., 2011). It should be noted that 'mean kurtosis', $\overline{W}$, calculated using axisymmetric fitting is equivalent to the 'mean kurtosis tensor' metric calculated when fitting the full kurtosis tensor. When reducing the number of diffusion directions for both fitting methods in HCP data, the reduced direction set that was most uniformly distributed along a sphere based on electrostatic repulsion was selected (Jones et al., 1999). It should be noted that the 10 directions used for the HCP data are not the same as the scheme outlined in Table 1. All b=0 images were used when the number of directions was reduced.

#### *2.5.2. Variation of principal eigenvector with frequency*

To determine the validity of using data from all OGSE frequencies to fit the diffusion tensor for the purpose of finding the symmetric diffusion axis, the variation of principal eigenvector with frequency was examined. A mask was manually drawn around the brain and the central angle between the 3-dimensional principal eigenvectors was found via the great-circle distance between them.

### *2.5.3. Methods to calculate the symmetric axis of diffusion*

We hypothesize that the symmetric axis of diffusion depends negligibly on b-value and OGSE frequency and, accordingly, it would be most robust to include data from all frequencies and b-values together when computing the axis of symmetry. To test this hypothesis, maps were computed with three different ways of determining the symmetric axis: (1) AFAB – Using data from all frequencies and all b-values to compute a single diffusion tensor in each voxel, with data from each frequency using the same symmetric axis for parameter fitting. (2) SFAB – Using data from all b-values to compute a diffusion tensor in each voxel for each frequency separately, with parameter fitting for each frequency using a different symmetric axis within each voxel (i.e., the entire fitting process is done independently for each frequency). (3) SFLB – Using data from only the low b-value shell (1000 s/mm$^2$) to compute a diffusion tensor in each voxel for each frequency separately, with subsequent parameter fitting for each frequency using data from all b-values.

To quantitatively compare these three approaches, region-of-interest (ROI) measurements were performed in the corpus callosum and hippocampus in mice. ROI masks were generated from the Turone Mouse Brain Atlas (Klein et al., 2009). First, the atlas was registered to a single 'chosen T2' volume from one of the scanning sessions using affine and symmetric diffeomorphic transforms with ANTs software (Avants et al., 2011). Similarly, the 'chosen T2' volume was registered into each subject's dMRI native space. The outputted deformation fields and affine transforms were then used to bring both ROIs into the native space of each subject. Masks were visually inspected to ensure good registration quality.

### *2.5.4. Comparing spatial regularization vs Gaussian smoothing*

To qualitatively compare noise levels in kurtosis maps using Gaussian smoothing and our spatial regularization algorithm, we chose to use $K_{\perp}$ maps as they showed the best contrast between grey and white matter. We implemented our regularization algorithm (during fitting) and Gaussian smoothing (prior to fitting) on diffusion-weighted images at various strengths which were chosen so that at each level both maps had an approximately equal visual appearance of noisy voxels. Comparisons between both methods were assessed qualitatively in both humans and mice and quantitatively in mice using the ROI analysis procedure outlined above.

## 3. Results

### 3.1. *Conventional kurtosis tensor vs. axisymmetric DKI fitting*

Figure 1 shows human data comparing the two fitting methods at varying numbers of diffusion directions, with no spatial regularization (i.e., $\gamma_{DT} = \gamma_{DK} = 0$). Both methods provide qualitatively similar maps, although maps computed using kurtosis tensor fitting have more noise. This noise reduction when using axisymmetric fitting is most apparent in $K_{\perp}$ maps, which becomes more discernible when the number of direction is reduced. $K_{||}$ maps show that axisymmetric fitting provides maps with higher contrast between white and grey matter whereas this contrast is difficult to discern when using tensor fitting. With fewer directions, the grey/white matter contrast becomes even worse for tensor fitting. These same differences are also evident in Figure 2 which shows mouse data comparing the two fitting methods with a traditional 40-direction scheme. Additionally, the efficient 10-direction scheme with averaging produces parameter maps with improved contrast and less noise as compared to maps acquired with more encoding directions, which show over-estimation of kurtosis parameters and noisy $\overline{D}$ and FA towards the bottom of the brain where there is low signal-to-noise ratio (SNR).

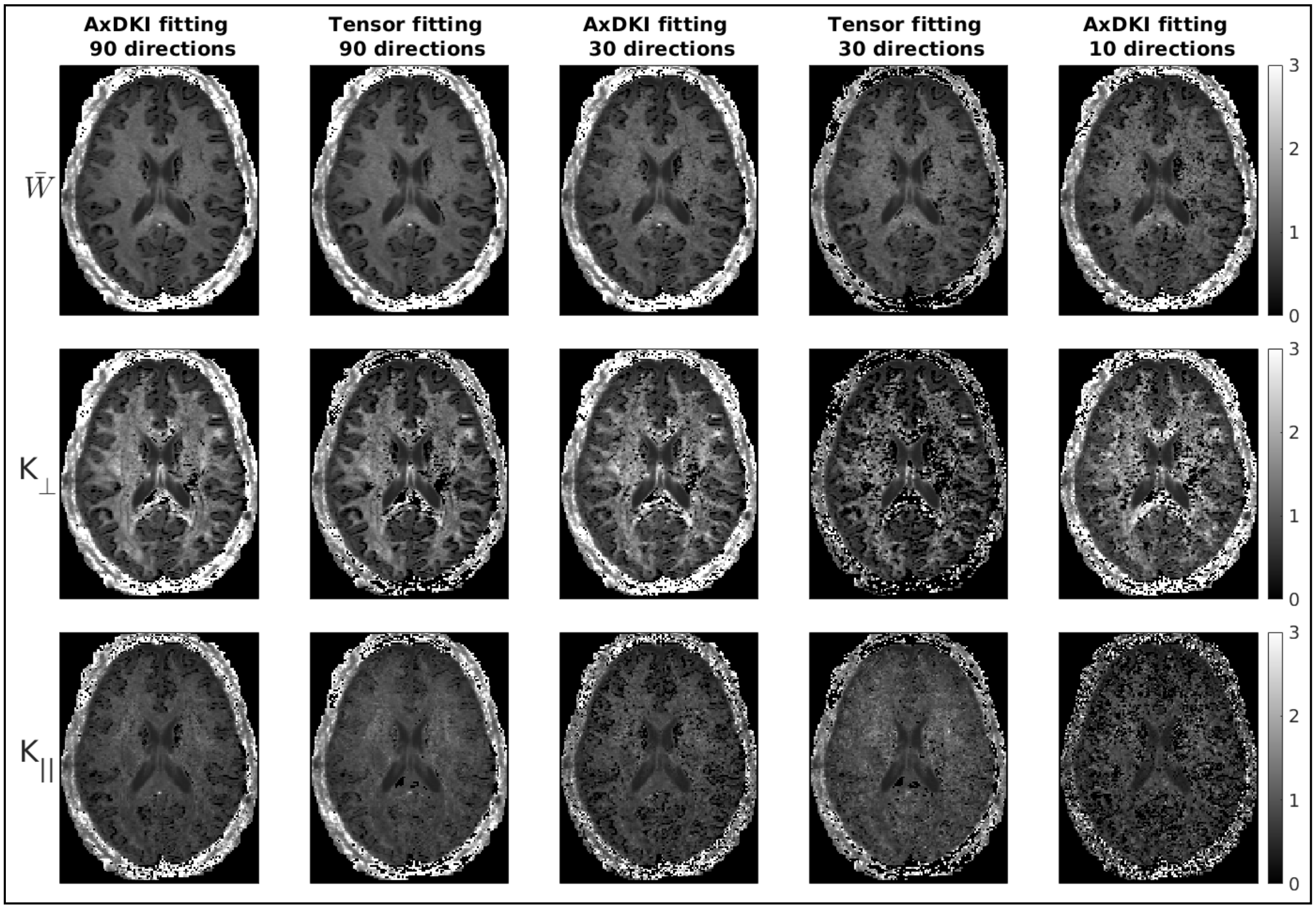

***Figure 1.*** *Comparison of kurtosis maps when using conventional kurtosis tensor fitting vs axisymmetric DKI (AxDKI) fitting at various numbers of diffusion directions. It should be noted that data shown with 10 directions is not the same scheme as outlined in Table 1.*

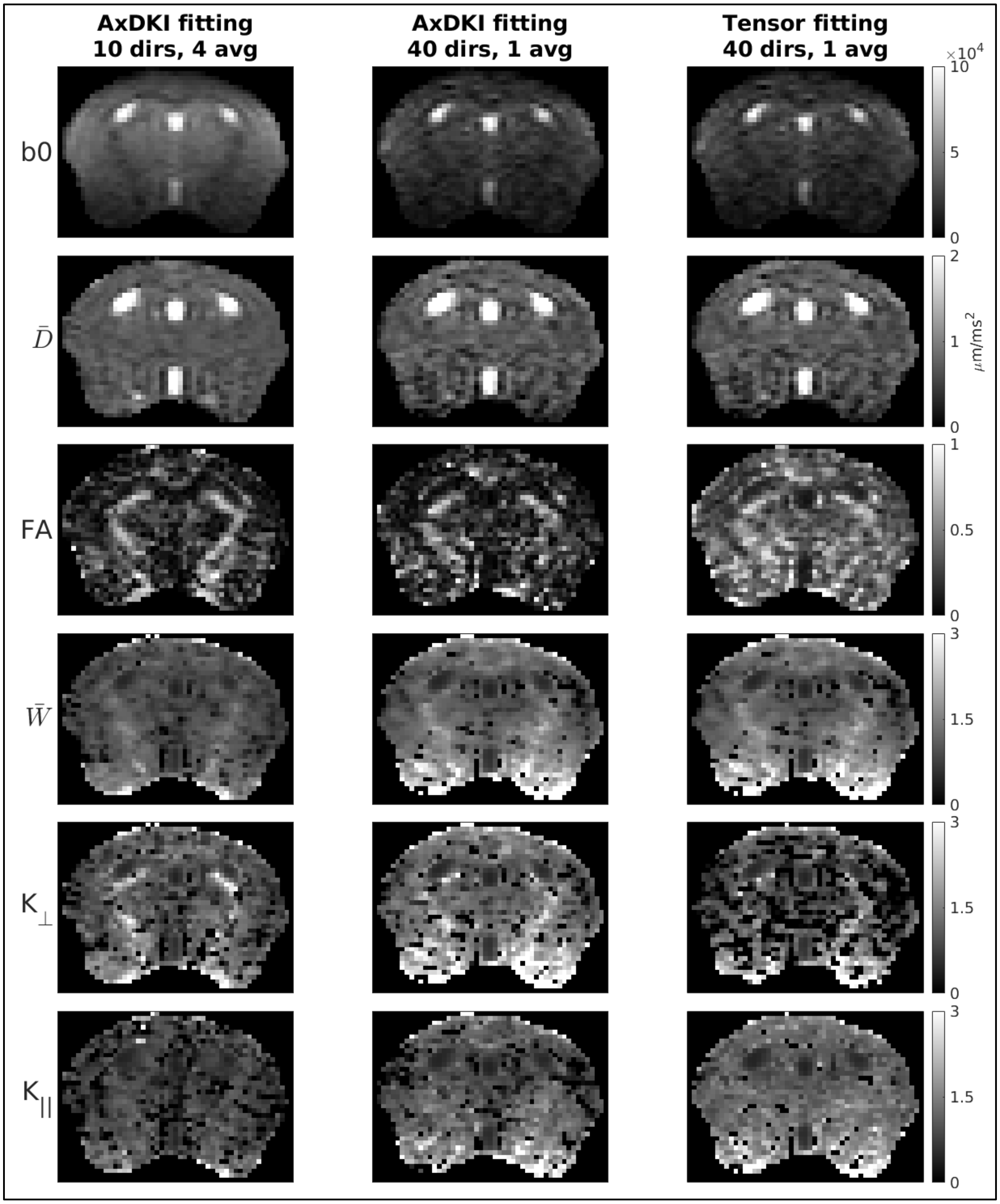

***Figure 2***. *Comparison of b=0 images, and diffusion tensor and kurtosis parameter maps computed with the 10-direction scheme with averaging with a traditional 40-direction scheme using axisymmetric DKI (AxDKI) fitting, and axisymmetric fitting vs kurtosis tensor fitting with the 40-direction scheme in a mouse. Data shown is from the PGSE (i.e., 0 Hz) acquisition.*

*3.2. Variation of principal direction of diffusion with frequency*

Figure 3a) shows a histogram of the agreement of the principal eigenvector across examined frequencies when using data from all b-value shells for computation. Good agreement across frequencies can be seen, with the best agreement being in voxels with high FA values. This can also be seen qualitatively in Figure 3b), which shows principal diffusion direction maps for each frequency. Differences between frequencies are difficult to discern qualitatively, especially in white matter regions. Supplementary Figure 1 shows similar histograms and maps but demonstrates how this agreement changes when using data from only the low b-value shell (1000 $s/mm^2$) to calculate the diffusion tensor. The distribution shown in the histogram is more skewed away from collinearity and it is also qualitatively apparent in the principal diffusion direction maps that the principal eigenvector shows noise-like inconsistencies across frequencies.

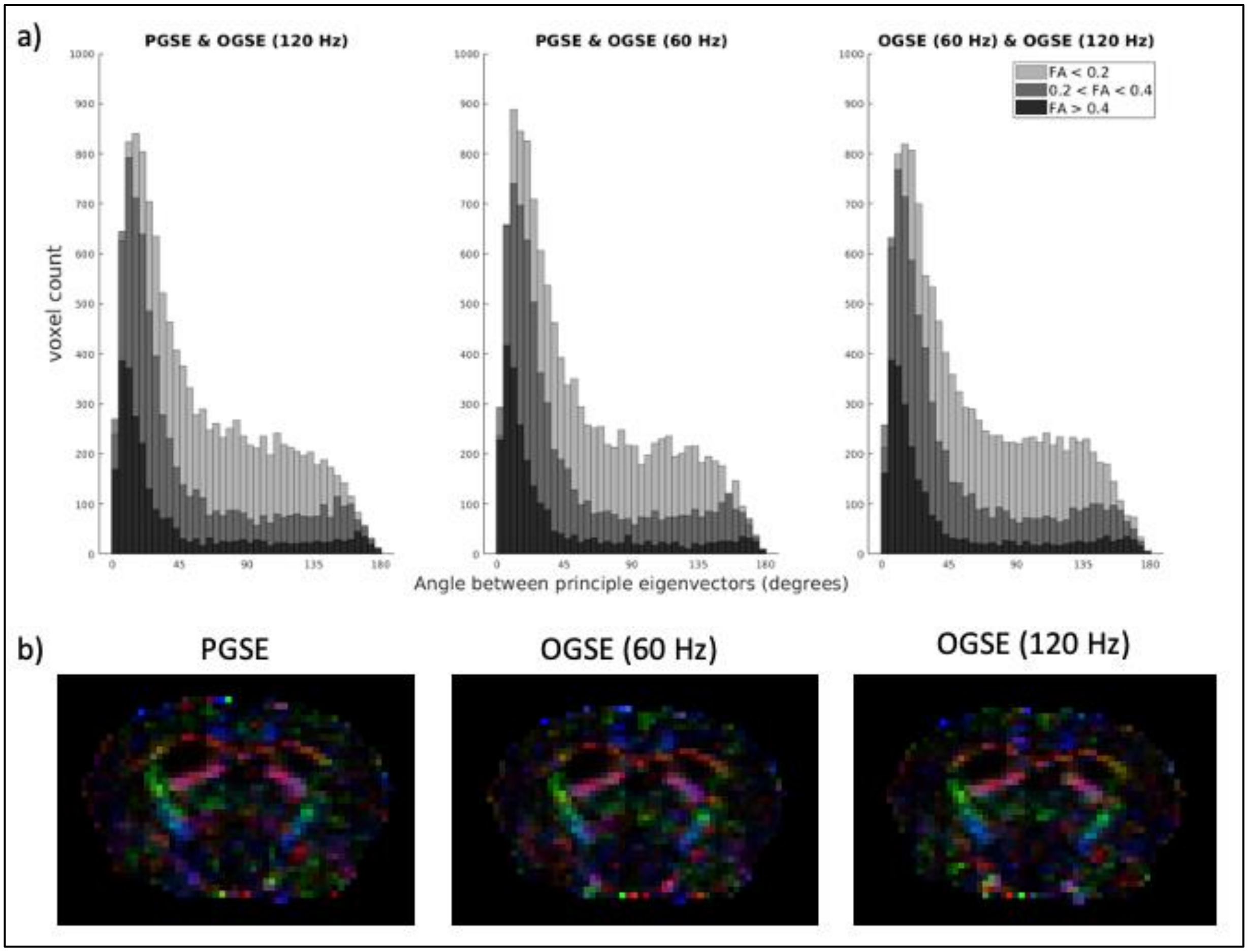

***Figure 3.*** *Examining the consistency of principal diffusion direction across PGSE and OGSE frequencies when using all b-value shells to calculate the diffusion tensor. a) shows histograms which quantify the central angle between principal eigenvectors computed at each frequency. b) shows principal diffusion direction maps for the corresponding acquisition above.*

*3.3. Comparing methods to calculate the symmetric axis of diffusion*

Figure 4 shows kurtosis and FA maps computed when using differing amounts of data to calculate the symmetric axis of diffusion via Eq. 8 with no regularization (i.e., $\gamma_{DT} = 0$). $\overline{W}$ maps appear qualitatively indistinguishable regardless of the method used to calculate the axis of symmetry. $K_{\perp}$ maps appear the least noisy when using data from AFAB to calculate the axis of symmetry, with quality degradation being most evident when using data from only the low b-value shell (SFLB) to designate the symmetric axis. $K_{\parallel}$ maps have an increased noise presence when not using data from all frequencies to determine the symmetric axis, and minimal contrast when using data from SFLB. FA maps also show the best image quality when using data from all frequencies and b-values to designate the symmetric axis.

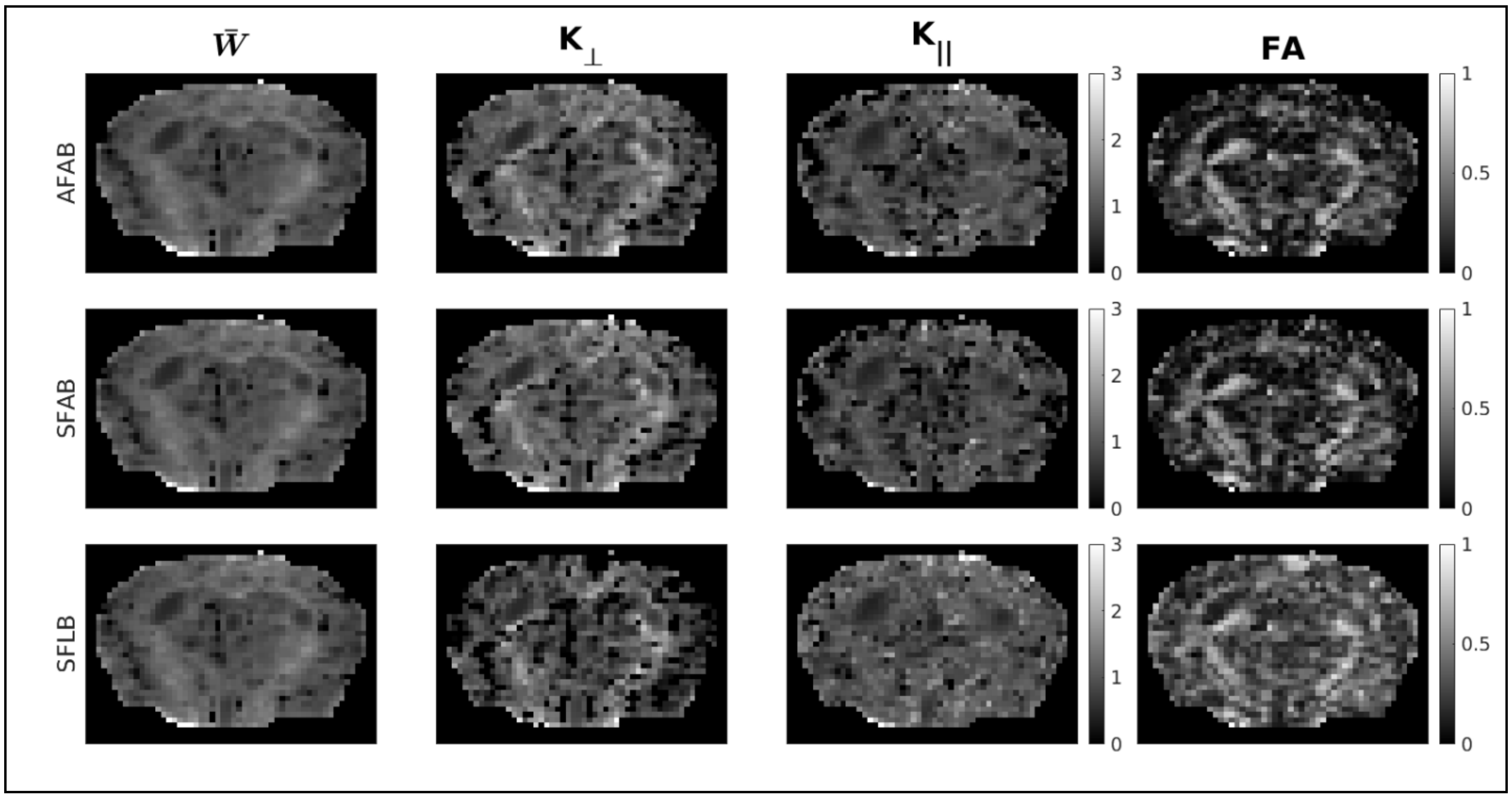

***Figure 4.*** *Comparison of kurtosis and FA maps when using various methods to calculate the diffusion tensor for axis of symmetry designation with no regularization (AFAB – using data from all frequencies and b-value shells, SFAB – using data from separate frequencies and all b-value shells, SFLB – using data from separate frequencies and only the low b-value shell). Data shown is from the PGSE (i.e., 0 Hz) acquisition.*

Figure 5 shows quantitative data when using different methods to calculate the axis of symmetry in the corpus callosum and hippocampus of a single mouse. As seen in Figure 4, $\overline{W}$ appears to be invariant to the axis of symmetry designation. Inclusion of some or all frequencies for diffusion tensor estimation resulted in similar quantitative values, with slightly decreased $K_{\parallel}$ when using data from SFAB. Importantly, the frequency dispersion of all kurtosis metrics remained relatively consistent no matter the method to calculate axis of symmetry. Similar results were observed in other subjects.

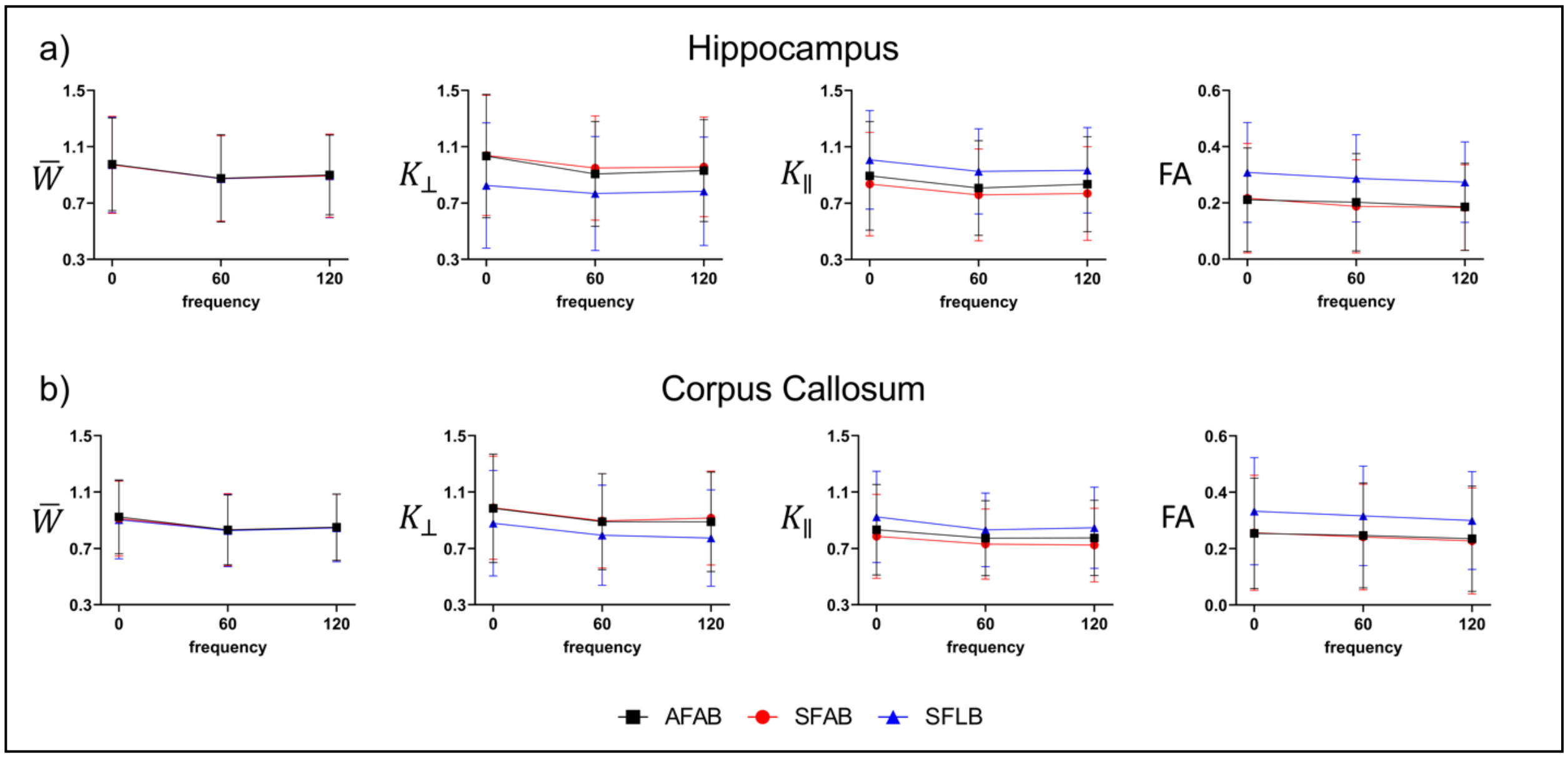

***Figure 5.*** *Quantitative data on how collapsing data from all frequencies and using all or only the low b-value shell to compute the axis of symmetry impacts kurtosis and FA values in a representative mouse (AFAB – using data from all frequencies and b-value shells, SFAB – using data from separate frequencies and all b-value shells, SFLB – using data from separate frequencies and only the low b-value shell). a) shows data from the hippocampus and b) shows data from the corpus callosum. All data was measured as mean +/- standard deviation within the same subject.*

### *3.4. Spatial regularization to control noise amplification*

Figure 6 shows kurtosis maps in mice with and without regularization. Regularization notably reduces the number of noisy voxels in all maps, while there is little-to-no blurring of the original contrast. This is especially evident in $K_\perp$ maps where there is high contrast between white and grey matter, which is conserved with regularization.

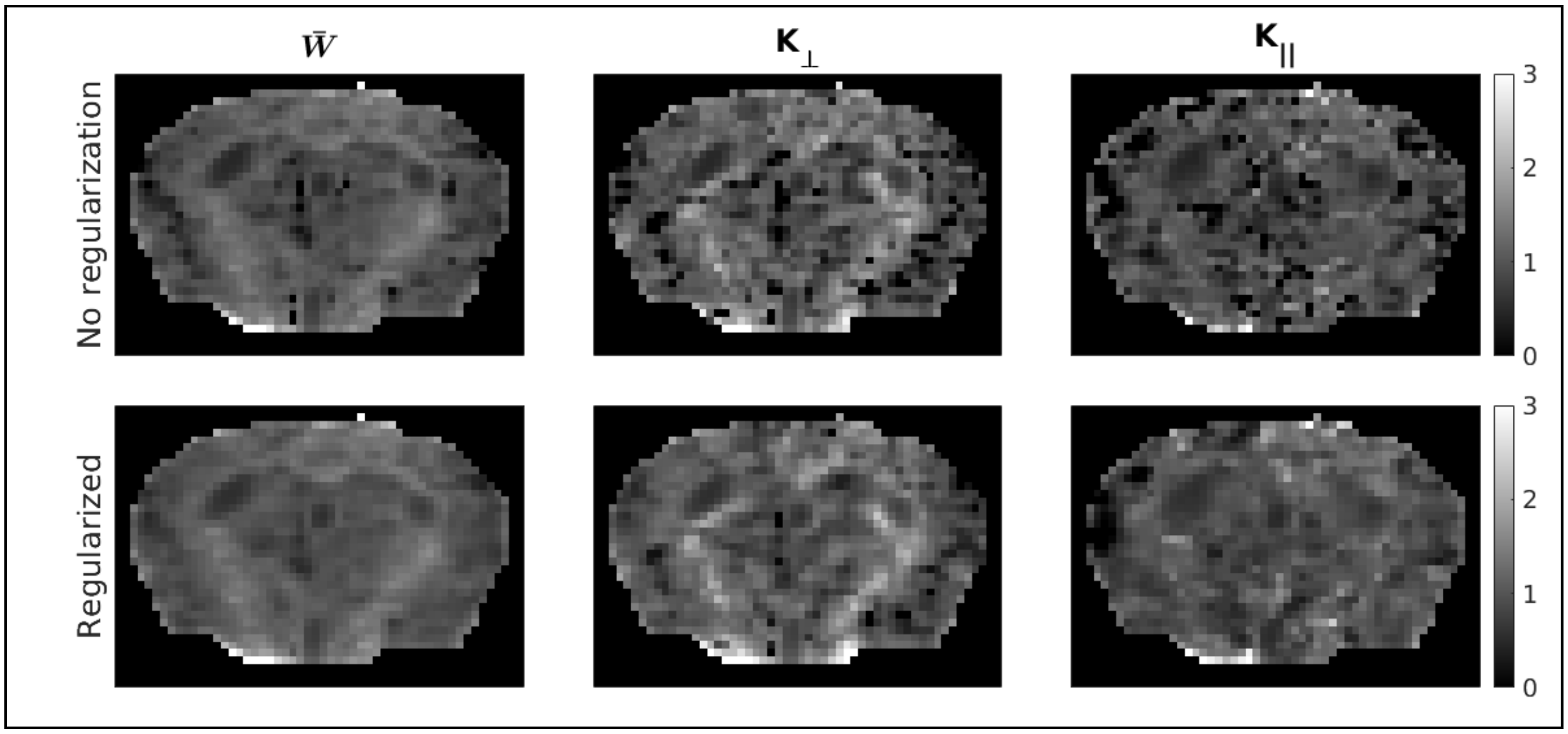

***Figure 6.*** *Comparison of kurtosis maps with and without implementation of our two-step regularization algorithm. The top row shows unregularized data and the bottom row shows data that produced high quality maps while avoiding over-*regularization ($\gamma_{DT} = 1.5, \gamma_{DK} = 0.225$). *Data shown is from the PGSE (i.e., 0 Hz) acquisition.*

Figure 7 shows a similar comparison of unregularized and regularized kurtosis maps in human data with only 10 diffusion directions, with unregularized maps computed using 30 and 90 directions shown for comparison. Similar to data shown in mice, regularization greatly reduces the number of noisy voxels in all maps while retaining original contrast. It is evident that the regularized data with only 10 directions produces parameter maps with similar or less noise than those acquired with 30 directions, at the expense of slightly reduced contrast at the borders between white and grey matter.

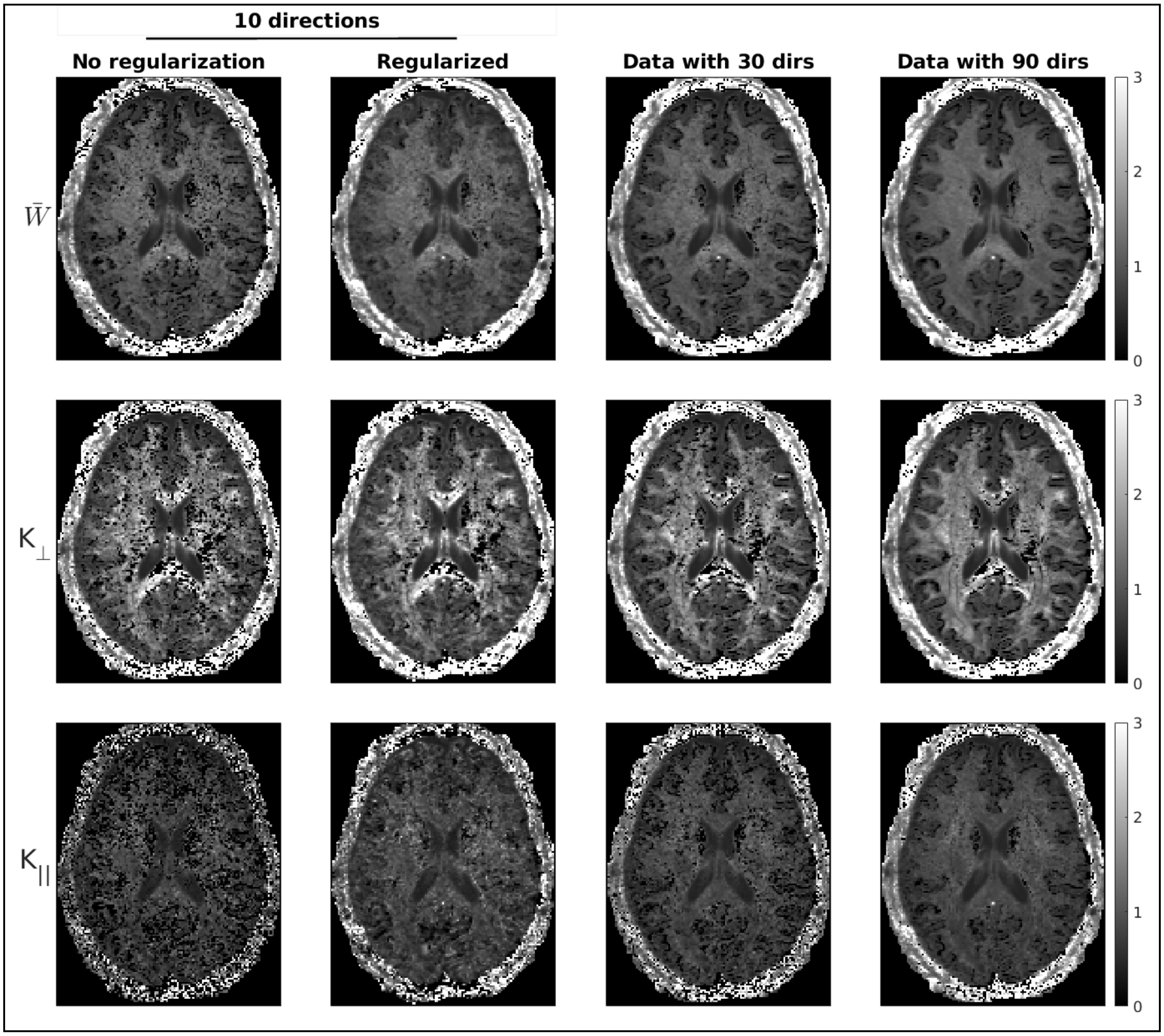

***Figure 7.*** *Comparison of kurtosis maps with* $(\gamma_{DT} = 1.5, \gamma_{DK} = 0.6)$ *and without regularization using 10 diffusion directions, and how regularized maps computed with only 10 directions compare with unregularized data computed using 30 and 90 directions.*

Figure 8 compares $K_{\perp}$ maps from a mouse with increasing levels of spatial regularization and Gaussian smoothing. Both methods reduce the number of noisy voxels, however, increasing the level of smoothing blurs the maps such that the contrast between white and grey matter becomes less apparent. In comparison, as the regularization weighting is increased, the number of noisy voxels also decreases but the original contrast is preserved. A similar comparison between regularization and smoothing is shown in human data in Supplementary Figure 2, which illustrates the same trends.

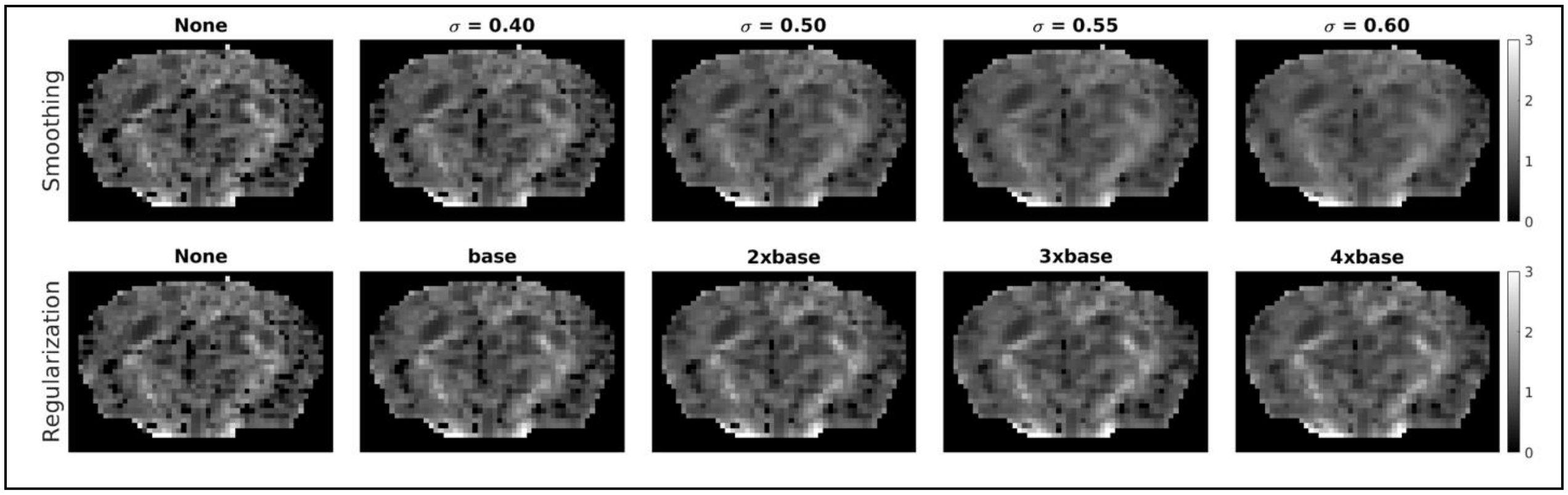

***Figure 8.*** *Comparison of $K_{\perp}$ maps with increasing levels of spatial regularization and Gaussian smoothing. $\sigma$ indicates the standard deviation of the Gaussian kernel used for smoothing on the diffusion-weighted images prior to fitting, in units of voxels. Regularization weighting and $\sigma$ levels were chosen so that in each column there are approximately equal numbers of noisy voxels in each column. Data shown is from the PGSE (i.e., 0 Hz) acquisition.*

Figure 9a) and 9b) show how increasing levels of regularization and smoothing impact the inter-subject variability of $\overline{D}$ and kurtosis metrics. It is apparent that with increasing levels of regularization the mean value of kurtosis metrics increases, which is expected as the number of noisy "blackened" voxels decreases. For Gaussian smoothing there are similar trends, with a larger amount of smoothing increasing the mean values in the hippocampus. However, in the corpus callosum, the mean value for each kurtosis metric remains consistent with increased smoothing while the diffusivity increases. This is likely due to a competing effect whereby noisy voxels are being removed, but there is blurring with nearby cerebrospinal fluid which decreases the kurtosis and increases diffusivity within white matter. This data also depicts that both regularization and Gaussian smoothing have little-to-no effect on the frequency dispersion of kurtosis metrics and diffusivity, and the inter-subject variability remains consistent suggesting this is due to true variation between subjects.

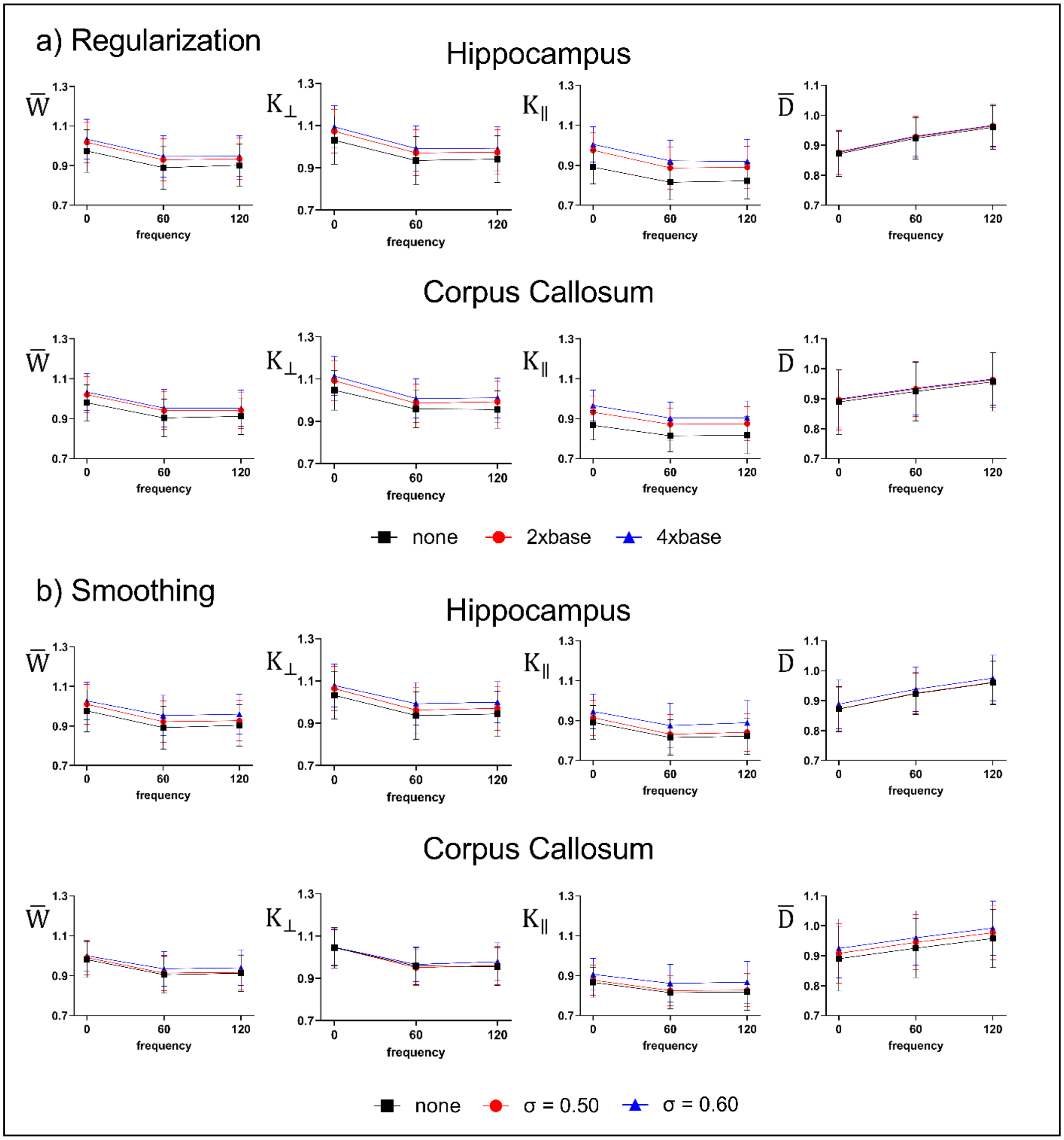

***Figure 9.*** *Examining how regularization and Gaussian smoothing impacts the inter-subject variability of kurtosis and diffusivity metrics. a) shows how regularization impacts the between subjects (n=8) mean and variability within the hippocampus and corpus callosum (base:* $\gamma_{DT} = 0.5, \gamma_{DK} = 0.075$*). b) shows how smoothing impacts the between subjects mean and variability within each region. Data shown as mean +/- standard deviation across subjects.*

## 4. Discussion

In this study, we investigate a method to compute frequency-dependent directional kurtosis maps, which was used in combination with a 10-direction scheme that has twice the efficiency of traditional schemes in generating b-value. Axisymmetric DKI fitting was shown to provide higher quality kurtosis maps as compared to conventional kurtosis tensor fitting, and the 10-direction scheme with multiple averages was shown to reduce erroneous kurtosis estimation as compared to traditional schemes with a larger number of directions due to the shorter TE enabled by the efficient direction scheme. The principal axis of diffusion did not vary appreciably between the frequencies explored, and we illustrated that using data from all frequencies and b-values for the purpose of determining the symmetric diffusion axis improved parameter map quality. Finally, as kurtosis fitting is susceptible to noise amplification, a two-step spatial regularization algorithm was presented, which reduces noise while preserving contrast and we showed the advantages of using regularization as opposed to conventional Gaussian smoothing for controlling noise amplification in DKI maps.

Axisymmetric DKI fitting was shown to produce higher quality kurtosis maps as compared to tensor fitting, becoming more evident when using fewer encoding directions (Figures 1 and 2). These results agree with two recent simulation studies which found comparable or even slightly improved performance of axisymmetric DKI compared to tensor fitting in terms of parameter estimation (Jespersen, 2018; Oeschger et al., 2023). Jespersen (2018) also showed that axisymmetric DKI generates somewhat more accurate $K_{\parallel}$ and $K_{\perp}$ estimation compared to tensor fitting in cases of low SNR, which agrees with our findings of greater increases in $K_{\parallel}$ and $K_{\perp}$ map quality when using axisymmetric DKI with fewer encoding directions. Our finding of $K_{\parallel}$ contrast being dominated by noise in tensor fitting agrees with Oeschger et al. (2023) who found axisymmetric DKI is particularly advantageous over tensor fitting in terms of $K_{\parallel}$ estimation. The improved performance of axisymmetric DKI is likely due to the reduced parameter space (8 compared to 22) used compared with tensor fitting, which decreases noise propagation during the fitting process.

The challenge in generating large b-values when using oscillating gradient encoding is evident in both rodent (Aggarwal et al., 2020) and human (Borsos et al., 2023) DKI where tetrahedral diffusion encoding was required to collect data at 140 Hz in mice and 23 Hz in humans. Although slightly less efficient in generating b-value, the 10-direction scheme was shown to generate b-values up to ~2500 s/mm$^2$ at frequencies up to 120 Hz with typical pre-clinical gradient strengths. While this encoding scheme is not optimized for uniform directional sampling, the trade-off is a large reduction in TE. Data collected with our efficient 10-direction scheme achieved a b-value of 2500 s/mm$^2$ at 120 Hz with a TE of 35.5 ms, while when a 40-direction scheme was used it required a TE of 52 ms due to the increased diffusion gradient duration required to achieve the same b-value. Despite the reduced number of encoding directions compared to a traditional DKI scheme with ~30-40 directions (Fukunaga et al., 2013; Tabesh et al., 2011), SNR can be recovered by using signal averaging as was done here. We show in Figure 2 that at the same scan time, an efficient encoding scheme with averaging produces maps with enhanced contrast in comparison to a higher direction scheme where there is visible over-estimation of kurtosis metrics in regions with low SNR (Zhang et al., 2021) (bottom of the brain as we use a surface coil placed on the top of the head) due to the increased TE. These results compliment the findings by Lebel et al. (2012), which showed that a 6-direction protocol with 5 averages gives comparable robustness in DTI metrics as compared to a 30-direction protocol even when the greater b-value efficiency of 6 directions is not exploited and TE's are held constant.

The most important step in the axisymmetric fitting method is the determination of a symmetric diffusion axis in each voxel, which is acquired by fitting the diffusion tensor and using the primary eigenvector as the symmetric axis. Results showed good agreement of the primary eigenvector across the frequencies explored (0 to 120 Hz), with the best agreement occurring in regions with high FA values and when using all b-values for computation (Figure 3 and Supplementary Figure 1). Using frequencies of 0 to 150 Hz, Aggarwal et al. (2012) found that the primary diffusion direction in voxels with relatively high FA had no apparent frequency-dependent change. The frequencies explored in this study correspond to hindered-diffusion, specifically average molecular displacements of ~3.5-8 μm as determined via the Einstein-Smoluchowski relation (Morozov et al., 2020). As this displacement range is larger than the typical diameters of axons in the mouse brain (< 1 μm) (Basu et al., 2023; West et al., 2015), the water molecules will reach the axonal membrane at all frequencies and therefore the primary diffusion direction should not vary between frequencies. This is seen as voxels with high FA values have better agreement between the symmetric axis across frequencies. In grey matter which is predominated by neuron, astrocyte, and microglia cell bodies with typical diameters of 5-10 microns (Bushong et al., 2002; Kozlowski & Weimer, 2012; Oberheim et al., 2009), it is possible that the primary diffusion direction could change between the frequencies explored. Nevertheless, in spherically shaped objects the diffusion tensor eigenvalues are approximately equal and the primary eigenvector is arbitrary and susceptible to noise.

The kurtosis parameter maps presented in Figure 4 show the least amount of noise when using data from all frequencies and b-values to calculate the symmetric axis of diffusion. This is expected as the primary diffusion direction is unlikely to change with frequency (as outlined above) as well as b-value (Tournier et al., 2013), therefore, using data from all frequencies and b-values will reduce the susceptibility of the diffusion tensor calculation to noise and increase its robustness. It was shown both qualitatively and quantitatively that $\overline{W}$ is invariant to the axis of symmetry designation, whereas $K_{\perp}$ and $K_{\parallel}$ map quality is dependent on an accurate designation of symmetric axis in agreement with previous findings (Hansen et al., 2016). Importantly, the frequency dispersion of kurtosis metrics remains consistent regardless of the method used to designate the axis of symmetry.

The implementation of the two-step regularization algorithm showed to be advantageous in reducing noise amplification in kurtosis fitting while preserving contrast in both mice and humans (Figures 6 and 7). Regularization remains underutilized in dMRI research despite showing improved image quality in simultaneous multi-slice dMRI data (Haldar et al., 2020), compressed sensing (Mani et al., 2021; Varela-Mattatall et al., 2023), and estimation of fiber orientation distribution functions (McGraw et al., 2009; Sakaie & Lowe, 2007). Our results show that regularization is particularly advantageous in improving parameter map quality in DKI, which are often confounded by noise amplification. Importantly, although this two-step algorithm was presented for the computation of frequency-dependent kurtosis maps, it can also be implemented for data acquired at a single diffusion time.

While Gaussian smoothing is a widely accepted as a necessary pre-processing step for DKI (Jensen et al., 2005; Tax et al., 2022), a direct comparison between smoothing and the proposed regularization implementation was provided, showing improved map quality when using regularization in both mice and human data (Figure 8 and Supplementary Figure 2). This aligns with functional MRI research where smoothing was shown to lead to a loss of detail caused by blurring of activation regions beyond their true boundaries (Liu et al., 2010). Comparatively, spatial regularization has been shown to improve activity detection and fine details, given that smoothing introduced via regularization is much less severe than Gaussian smoothing (Casanova et al., 2009). When comparing the regularized parameter maps to those obtained with Gaussian smoothing, it is evident that regularization preserves image contrast much better than smoothing which causes blurring with adjacent structures and cerebrospinal fluid. Although not explored here, spatial regularization is likely to have advantages over Gaussian smoothing when conducting voxel-wise analyses, as smoothing can lead to large distortions in voxels which vary depending on their neighboring voxels. This is because spatial regularization preferentially smooths the largest errors more strongly (i.e., those with large spatial gradients), whereas Gaussian smoothing non-specifically smooths all voxels in an image equally. Finally, we showed that both regularization and Gaussian smoothing have little impact on the dispersion of kurtosis metrics with frequency and inter-subject variability, which is of great importance when examining the frequency-dependence of metrics.

This study is not without its limitations. As alluded to previously, based on the sizes of microstructural barriers within tissue, it is not expected that the primary diffusion direction would change across the frequencies explored in this study. However, this assumption of unity of the principal diffusion axis across frequencies may not always be true, especially when using very high OGSE frequencies. Although the workflow generates high-quality kurtosis maps in mice, further investigation of the proposed method should be investigated in humans. Based on recent work by Dai et al. (2023) who performed frequency-dependent DKI in humans with b-values of 2000 s/mm$^2$ and frequencies up to 47.5 Hz with the MAGNUS high performance gradient system, using the more efficient 10-direction scheme and our fitting algorithm could allow for increased map robustness and higher b-values and frequencies to be examined in humans. Additionally, incorporation of spiral trajectories would allow for increased b-value for a given TE as compared to standard Cartesian trajectories (Michael et al., 2022). Future work may also explore how different types of regularization (i.e., joint regularization (Bredies et al., 2010)) can be used to more effectively control noise amplification in kurtosis maps. The $\ell_2$-norm based regularization method was chosen here because it can be evaluated rapidly using the conjugate gradient method. The computation time was 7 seconds for a full brain mouse dataset and 64 seconds for human HCP dataset on a workstation with a Intel(R) Core(TM) i9-11900K (3.5 GHz) and an NVIDIA GeForce RTX 4090 (24 GB). Additionally, the analysis scheme presented could benefit from automatic selection of the regularization weighting strength (Varela-Mattatall et al., 2021).

## 5. Conclusions

In conclusion, we presented a workflow to generate robust frequency-dependent kurtosis maps in mice. The 10-direction encoding scheme presented is twice as efficient in generating b-value compared to traditional schemes, and axisymmetric DKI fitting was shown to provide kurtosis maps with less noise than conventional tensor fitting and has reduced dataset requirements that enables fitting with a 10-direction scheme. We showed that using this 10-direction scheme with multiple averages is advantageous in terms of kurtosis parameter estimation as opposed to traditional schemes with ~30-40 encoding directions. Furthermore, taking advantage of degeneracies across frequencies and implementing a two-step regularization algorithm was shown to decrease noise amplification while maintaining image contrast. While the interpretation of changes in frequency-dependent kurtosis parameters remain unclear, this workflow will allow further study of how these changes relate to tissue microstructure and may highlight the frequency-dependence of DKI metrics as useful biomarkers in the study of various conditions.

## 6. Acknowledgements

The authors would like to thank Amr Eed, Alex Li, and Miranda Bellyou for technical assistance with MRI scans and animal set-up.

This research was supported by the Natural Sciences and Engineering Research Council of Canada: Canada Graduate Scholarships—Master's Program (NSERC-CGS M), Canada Research Chairs (950-231993), Canada First Research Excellence Fund to BrainsCAN, and the US Department of Defense under congress-directed medical research program (CDMRP), Peer Reviewed Alzheimer's Research Program (PRARP) by award# W81XWH-20-1-0323.

## 7. Data and code availability statement

Code for our implementation of axisymmetric DKI fitting with optional spatial regularization is available at https://gitlab.com/cfmm/matlab/matmri. Human data used in this study is available from the Human Connectome Project at https://www.humanconnectome.org/study/hcp-young-adult/document/1200-subjects-data-release, and mouse data is available upon reasonable request.

## 8. Author contributions

**Jake Hamilton**: Conceptualization, Data curation, Formal analysis, Investigation, Visualization, Writing – original draft, Writing – review & editing. **Kathy Xu**: Project administration, Resources, Writing – review & editing. **Arthur Brown**: Funding acquisition, Project administration, Resources, Supervision, Writing – review & editing. **Corey A. Baron**: Conceptualization, Funding acquisition, Resources, Software, Supervision, Writing – review & editing.

## 9. Declaration of competing interests

The authors declare no competing interests.

## 10. Supplementary Material

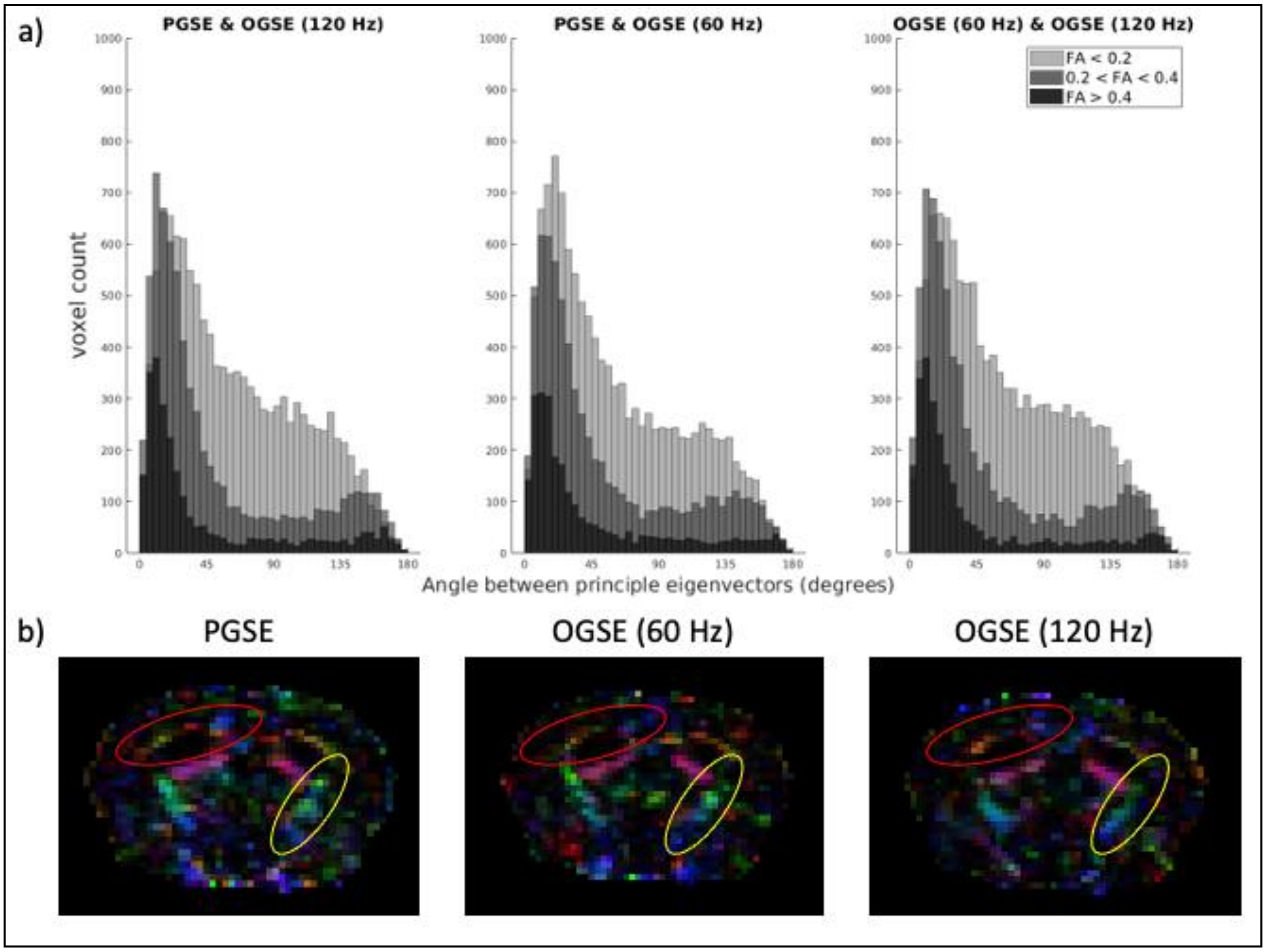

***Supplementary Figure 1***. *Examining the consistency of principal diffusion direction across PGSE and OGSE frequencies when using only the low b-value shell to calculate the diffusion tensor. a) shows histograms which quantify the central angle between principal eigenvectors computed at each frequency. b) shows principal diffusion direction maps for the corresponding acquisition above. The corpus callosum is outlined in red and the internal capsule in yellow.*

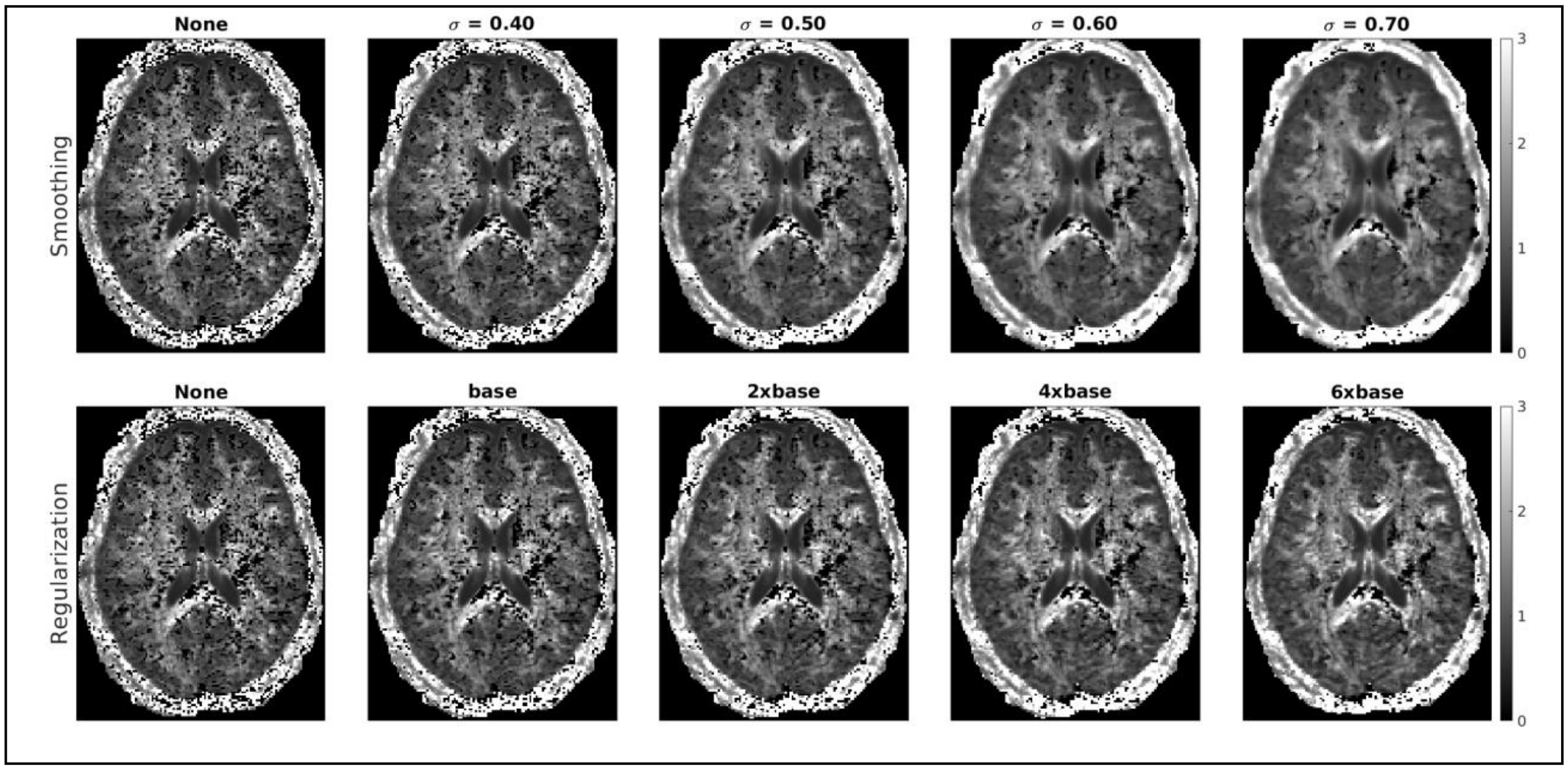

***Supplementary Figure 2.*** *Comparison of $K_{\perp}$ maps with increasing levels of spatial regularization and Gaussian smoothing. $\sigma$ indicates the standard deviation of the Gaussian kernel used for smoothing on the diffusion-weighted images prior to fitting, in units of voxels. Regularization weighting and $\sigma$ levels were chosen so that in each column there are approximately equal numbers of noisy voxels in each column.*